# Introducing Cashless Transaction Index based on the Effective Medium Approximation

Mikrajuddin Abdullah

Department of Physics, Bandung Institute of Technology

Jalan Ganesa 10 Bandung 40132, Indonesia

Email: mikrajuddin@gmail.com

**Abstract**

The effective medium approximation (EMA) method is commonly used to estimate the effective conductivity development in composites containing two types of materials: conductors and insulators. The effective conductivity is a global parameter that measures how easily the composite conducts electric current. Currently, financial transactions in society take place in cash or cashless, and, in the cashless transactions the money flows faster than in the cash transactions. Therefore, to provide a cashless grading of countries, we introduce a cashless transaction index (CTI) which is calculated using the EMA method in which individuals who make cash transactions are analogous to the insulator element in the composite and individuals who make cash transactions are analogous to the conductor element. We define the CTI as the logarithmic of the effective "conductivity" of a country's transactions. We also introduce the time dependent equation for the cashless share. Estimates from the proposed model can explain well the data in the last few years.

# 1. Introduction

Millions of financial transactions, either in cash and cashless [1], generate large scale patterns (network) in the money flow [2,3]. The money flow a financial network [4] resembles the current flow in an electrical network so it may raise a question whether the equation describing the current flow in the electrical networks is adoptable for explaining the money flow in its corresponding network. An example of electrical network is a composite of two types of elements having different conductivities: a conductor and an insulator. The effective (global) conductivity development in such a composite was commonly calculated using the effective medium approximation (EMA) [5-9]. The EMA has also been applied in economics, specifically to predict the result of interaction between sellers and buyers (trade agents) [10].

The quantity belongs to the financial network that mimics the volume fraction of the conductor elemwnt in the composite is the cashless share. The Nordic countries especially Norway and Sweden have a high cash share, where the cash is less than 2.5% of total money supply [11,12] and those conutries have been seen as heading towards cashless societies. In general, the European economies have tended to move to a cashless society [12]. In Asia-Pacific, the cashless transaction volume has increased by 109% from 2020 to 2025 and then by 76% from 2025 to 2030, followed by Africa (78%, 64%) and Europe (64%, 39%). Latin America comes next (52%, 48%), and the US and Canada will have the least rapid growth (43%, 35%) [13]. Based on these evidences, it is likely desired to introduce a parameter to grade the cashless level of countries. To the best of our knowledge, there are rare reports related to such a grading. The present grading has been introduced by Mastercard Advisor by introducitng four grading: inception (the lowest), transitioning, tipping point, and nearly cashless (the highest) [14].

In this paper, we propose a parameter called cashless transaction index (CTI) to measure the cashless grading countries calculated using the EMA method. We use this method by considering the cashless grade must be a global parameter, similar to the effective electrical conductivity. We also propose a model to explain the growth of the cashless share.

## 2. Modeling

### 2.1 Cashless transaction index

A society (country) is treated as a "composite" of two types of individuals who make either cash or cashless transaction (see **Fig. 1**). The country's resident space is divided into identical cells where each cell is filled by one individual. The cashless transactions have a number of advantages, such as more convenience, more secure, gain many discounts, can pay anywhere [15], reduce floating costs, achieved instantaneously [16], and increase economic growth [17] to about 0.08% of GDP in developing countries [18] so that individuals who carry out cashless transactions can be analogized as the conductor in the composite. On the other hand, he cash transactions take time to get at, riskier to carry, and by most estimates it costs society as much as 1.5% of GDP [14], which implies that money will experience flow bottlenecks in the network. Thus, individuals who carry out the cash transactions are identical to insulating elements (**Fig. 2**).

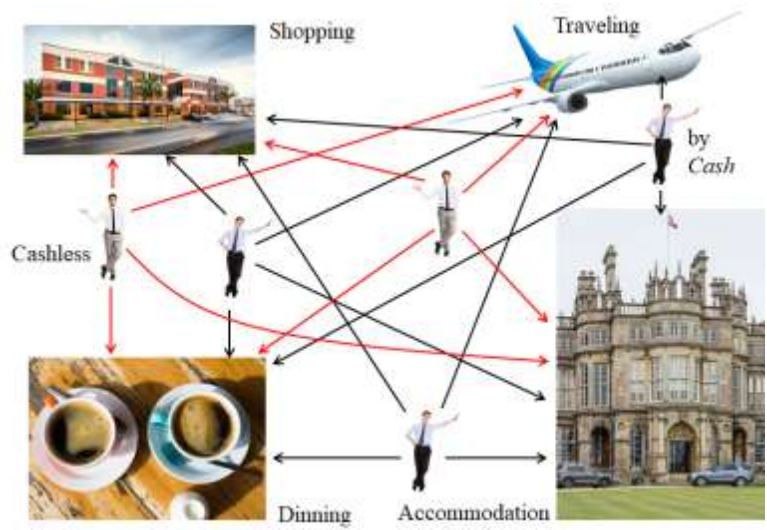

**Figure 1** Illustration of cash and cashless transactions. In general, individuals are divided into two groups: groups that carry out cash transactions and cashless transactions. Each individual can access the transaction center for shopping, traveling, dinning, accommodation, and so on. Individuals linked by red arrows make cashless transactions and those marked by black arrows make cash transactions [image source: free online pictures from Microsoft Power Point].

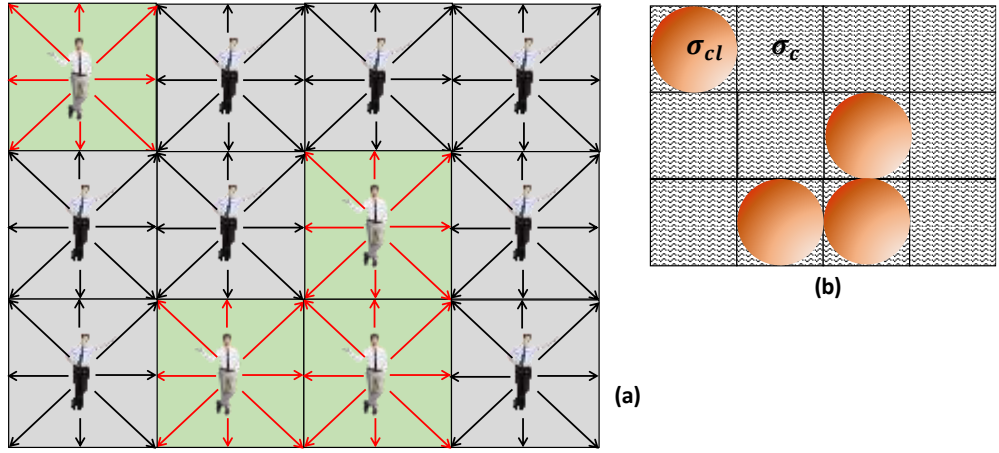

**Figure 2** (a) A society divided into identical cells. Each cell is filled by individuals who make transactions by cash and individuals who carry out cashless transactions. (b) The analogy for Figure (a) is a composite formed by conducting particles dispersed into an insulating matrix.

In this paper, the effective (global) conductivity, $\sigma_e$, is calculated using the equation [5-9] $\sum_i f(\sigma_i)(\sigma_i - \sigma_e)/(\sigma_i + (z/2 - 1)\sigma_e) = 0$ which can be separated as

$$\sum_i f(\sigma_{c,i}) \frac{\sigma_{c,i} - \sigma_e}{\sigma_{c,i} + (z/2-1)\sigma_e} + \sum_j f(\sigma_{cl,j}) \frac{\sigma_{cl,j} - \sigma_e}{\sigma_{cl,j} + (z/2-1)\sigma_e} = 0 \qquad (1)$$

where $\sigma_{c,i}$ is the cash transaction conductivity of the i-th element, $\sigma_{cl,j}$ is the cashless transaction conductivity of the j-th element, $f(\sigma_{c,i})$ is the cash transaction fraction of the i-th element, $f(\sigma_{cl,j})$ is the cashles transaction fraction of the j-th element, and $\sum_i f(\sigma_{c,i}) + \sum_j f(\sigma_{cl,j}) = 1$, and $z$ is the coordination number (number of nearest neighbors). If we assume $\sigma_{c,i} \cong \sigma_c$ and $\sigma_{cl,j} \cong \sigma_{cl}$ for all i and j, Eq. (1) becomes

$$p \frac{\sigma_{cl} - \sigma_e}{\sigma_{cl} + (z/2-1)\sigma_e} + q \frac{\sigma_c - \sigma_e}{\sigma_c + (z/2-1)\sigma_e} = 0 \qquad (2)$$

where $p = \sum_j f(\sigma_{cl,j})$ and $q = \sum_i f(\sigma_{c,i}) = 1 - p$ are cashless shares and cash shares, respectively. The $p$ and $q$ are time dependent so as $\sigma_e$. Equation (2) is nothing that a quadratic equation, $A\sigma_e^2 + B\sigma_e + C = 0$, where $A = z/2 - 1$, $B = -(p\sigma_{cl}A - p\sigma_c + q\sigma_c A - q\sigma_{cl})$, and $C = -\sigma_c \sigma_{cl}$.

At present we give $\sigma_c = 1$ and $\sigma_{cl} = 10$ although other values are allowed as long as $\sigma_{cl} > \sigma_c$. Once we have $\sigma_e$ on all $p$, the next question is how to define the CTI? We propose the definition for the CTI on a logarithmic scale as

$$CTI = 10 \frac{\log_{10}(\sigma_e)}{\log_{10}(\sigma_{cl})} \tag{3}$$

It is clear that, if $p = 0$, $\sigma_e = \sigma_c = 1$ so $CTI = 0$, and if $p = 1$, $\sigma_e = \sigma_{cl}$ so $CTI = 10$. In fact, many quantities are defined on a logarithmic scale such as the pH, the sound intensity level, the volcano explosion index [19], and others.

## 2.2 Cashless share growth

The cashless share increases with time. In the early stages, it increases slowly with time, then rises sharply, and finally rises slowly, mimicking a sigmoid curve. To produce such a change, we assume that the cashless share rate depends on the probability of contacts between cashless-cashless agents ($P_{cl-cl}$), cashless-cash agents ($P_{c-cl}$), and cash-cash agents ($P_{c-c}$). We propose the following equation $dp/dt = a_1 P_{cl-cl} + a_2 P_{c-cl} + a_3 P_{c-c}$, and with the hypothesis [20-22] $P_{cl-cl} = p^2, P_{c-c} = q^2 = (1-p)^2, P_{c-cl} = 2pq = 2p(1-p)$, and $a_1, a_2$, and $a_2$ are parameters that are indepemdet of $p$ or $q$, the cashless share rate becomes

$$\frac{dp}{dt} = a_1 p^2 + a_2 p(1-p) + a_3(1-p)^2 \tag{4}$$

The initial conditions are, $dp/dt \to 0$ if $p \to 0$ so that $a_3 = 0$ and $dp/dt \to 0$ if $p \to 1$, so that $a_1 = 0$. Thus Eq. (4) takes the simple form as

$$\frac{dp}{dt} = a_2 p(1-p) \tag{5}$$

If $a_2$ is a constant, Eq, (5) is a classical logistic equation which has the general form $dp/dt = rp(t)(1 - p(t)/K)$ [23,24] where for our case $K = 1$. The logistic equation is one of the models that describe the growth of a single species population with limited resources [25]. We also have limited resources because the total population size is almost constant (we ignore the additional number of adults allowed to transact over a period of several decades compared to the total number of adults). Liu and Wang [23] explain the stability of the solution of the logistic equation by introducing the grow rate is not a constant, but a random function of time with the transformation $a_2 \to a_2 + \sigma \dot{B}(t)$ where $B(t)$ is the standard Brownian motion and $\sigma^2$ represents the intensity of white noise. When describing bacterial growth kinetic, Egli proposed that the "kinetic constant" of growth does not have to be constant, but can be a function of time [26]. In this paper we also assume that in general $a_2$ is time dependent so that the general solution of Eq. (5) is

$$p(t) = \frac{1}{1+e^{-(\varphi-\mu)}} \tag{6}$$

where

$$\varphi = \int_0^t a_2(t)dt \tag{7}$$

and $\mu$ is a constatnt. We will try two kinds of parameter $a_2$ namely as a constant or a linear function of time.

Let us write Eq. (6) in the form $\ln(1/p(t) - 1) = -\varphi + \mu$. If $a_2$ is a constant, $\varphi = a_2 t$ and

$$\ln\left(\frac{1}{p(t)} - 1\right) = -a_2 t + \mu_1 \tag{8}$$

If $a_2(t) = a_{20} + bt$ we have

$$\ln\left(\frac{1}{p(t)} - 1\right) = -\frac{1}{2}bt^2 - a_{20}t + \mu_2 \tag{9}$$

## 3. Results and Discussion

### 3.1 Time-dependent CTI of selected countries

**Figure 4** is the result of calculations on the CTI of several countries using data from statista [27-38] between 2000 and 2020. Some countries have data over a long period of time, while others only have 5 years of data. Croatia, Japan, Slovakia, Slovenia, Italy, and Hungary still have low CTIs, which are below 4 until 2020. Portugal, Netherland, Finland, UK, Sweden, and Denmark are countries that have approached a cashless society [39]. Japan has only experienced an increase of about 150% in the last 10 years and is still at 2.6 in 2020. Until now, the cashless payment rate is still low in Japan, which is only about 20% in 2017 and the government plans to increase it to 40% in 2027 [40].

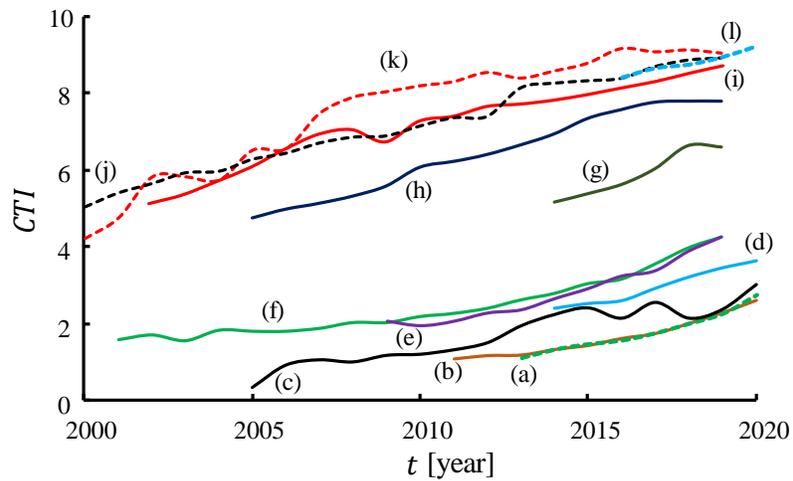

**Figure 4** The CTI values of various countries in the period 2000 – 2020: (a) Croatia [27],(b) Japan [28], (c) Slovakia [29], (d) Slovenia [30], (e) Italy [31], (f) Hungary [32], (g) Portugal [33], (h) Netherland [34], (i) Finland [35], (j) UK [36], (k) Sweden [37], and (l) Denmark [38].

Adopting the classification given by MasterCard Advisors [14], we call the first region having $0 \leq CTI \leq 2.5$, corresponds to $0 \leq p \leq 0.2864$ as *inception*. We call the second region, $2.5 \leq CTI \leq 5$ corresponds to $0.2864 \leq p \leq 0.5$ as *transitioning*. We call the third region, $5 \leq CTI \leq 7.5$ corresponds to $0.5 \leq p \leq 0.7136$ as *tipping point*. We call the fouth region, $7.5 \leq CTI \leq 10$ corresponds to $0.7136 \leq p \leq 1$ as *nearly cashless*. See **Fig. 5** as illustration.

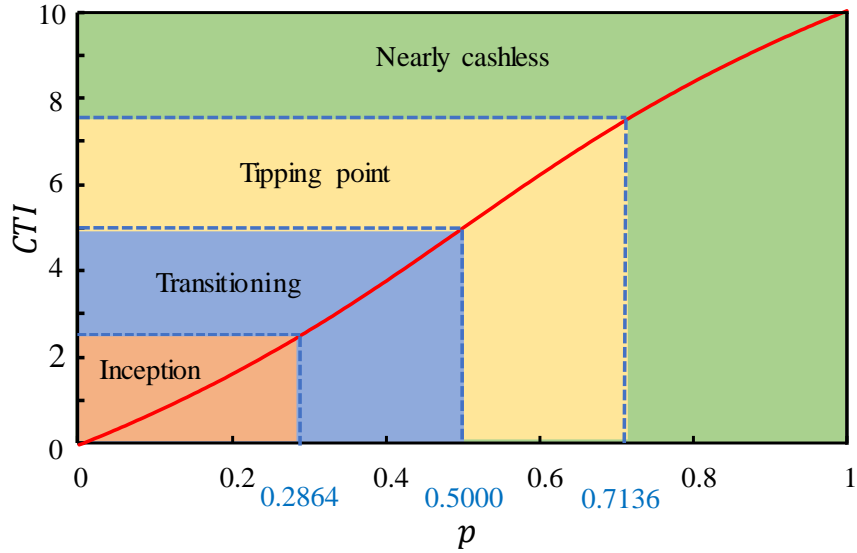

**Figure 5** CTI division into four regions: inception, transitioning, tipping point, and nearly cashless.

## 3.2 Time-dependent cashless share of selected countries

**Figure 6** is the data $\ln(1/p - 1)$ as a function of time for Finland, Sweden, United Kingdom, Netherland, Denmark, and Portugal. These countries have high cashless shares, can be considered as having implemented cashless activities for a long time. The data for these countries can be accurately fitted using $a_2$ as a constant (Eq. (8)). We have also tried fitting using $a_2$ as a linear function of time, but in general it results in a smaller $R^2$.

**Figure 7** is the data $\ln(1/p - 1)$ as a function of time for Slovenia, Italy, Japan, Croatia, Slovakia, and Hungary. These countries still have low cashless shares, mostly are only in the early stages of implementing cashless activities, later than the countries mentioned in **Fig. 6**. The data for all these countries can be accurately fitted using $a_2$ as a linear function of time (Eq. (9)). We also tried fitting using $a_2$ as a constant, but in general it results in a smaller $R^2$. Thus, we can expect when a country has just started the cashless activities, $\ln(1/p - 1)$ is a quadratic function of time, whereas when a country has been implementing cashless activities for a long time, $\ln(1/p - 1)$ is a linear function of time.

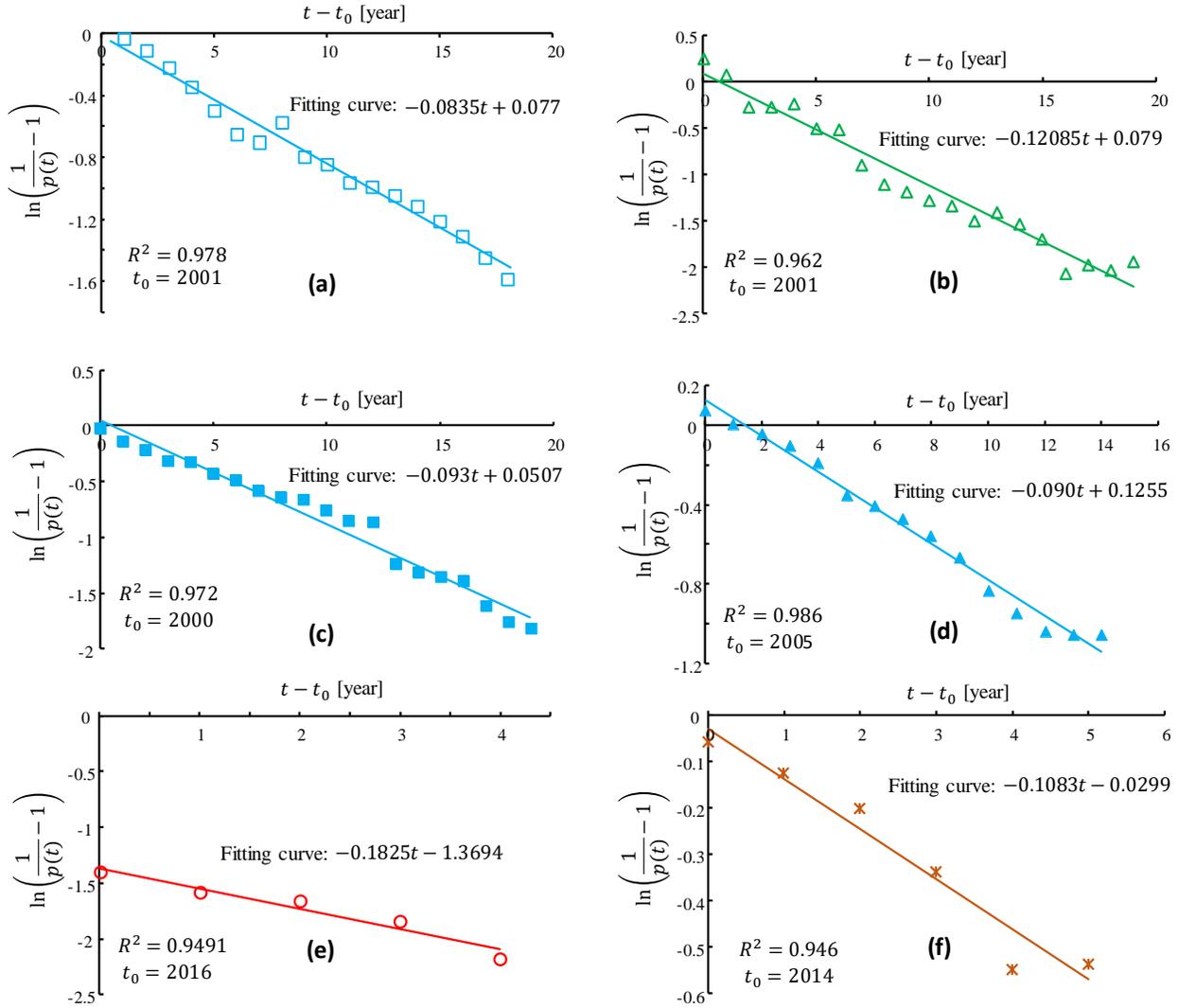

**Figure 6**. $\ln(1/p - 1)$ as a function of time for some countries: (a) Finland, (b) Sweden, (c) United Kingdom, (d) Netherland, (e) Denmark, and (f) Portugal. Symbols are data from (a) [35], (b) [37], (c) [36], (d) [34], (e) [38], and (f) [33]. Lines are linear fittings for each data set. We have used $t_0$ as the earliest year of the available data.

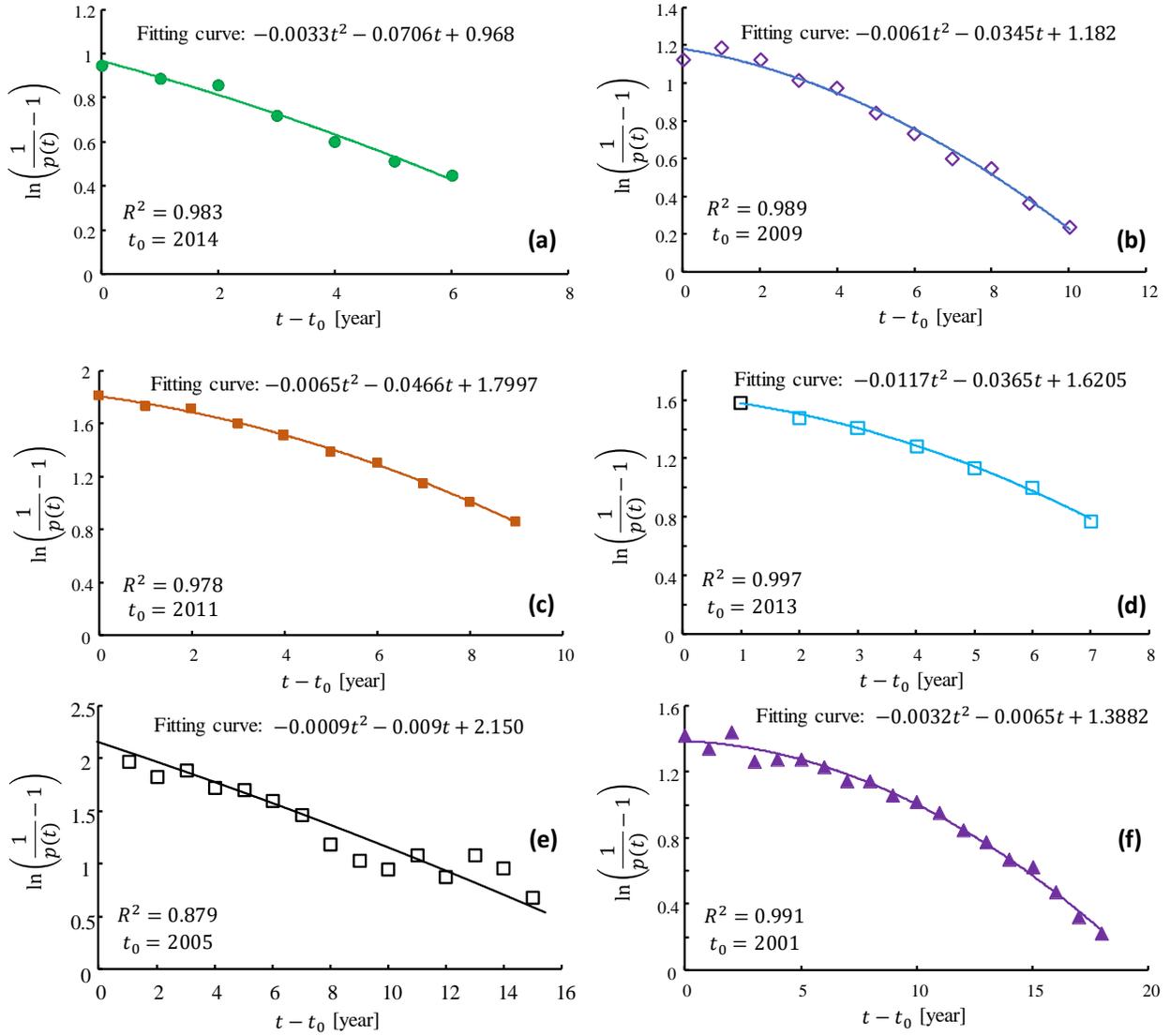

**Figure 7**. $\ln(1/p - 1)$ as a function of time for some countries: (a) Slovenia, (b) Italy, (c) Japan, (d) Croatia, (e) Slovakia, and (f) Hungary. Symbols are data from (a) [30], (b) [31], (c) [28], (d) [27], (e) [29], and (f) [32]. The curves are quadratic fittings for each data set. We have used $t_0$ as the earliest year of the available data.

In **Figs. 6** and **7** we have used $t_0$ as the earliest year of the available data so that $t_0$ is different for different countries. $t_0$ in **Fig. 6** is not the start time of cashless activity in that country. Cashless-based activities in those countries have actually been going on for quite a long time, which is indicated by the already high cashless share value in the early years of the data. Thus, the initial year of cashless activity should have occurred at time $t_0 - \Delta t_0$. In the next section we will estimate the value of $\Delta t_0$ for the countries in **Fig. 6**.

On the other hand, the countries in **Fig. 7** can be seen as countries that have just started to activate cashless activity (relative to the countries in **Fig. 6**). The earliest year from the data can be assumed to be the year when significant cashless activity started in these countries. Therefore, in these countries we approximate $\Delta t_0 \approx 0$. If we calculate from the start time of cashless activities for the countries in **Fig. 6**, Eq. (8) can be written as

$$\ln\left(\frac{1}{p(t)} - 1\right) = -a_2(t - \Delta t_0) + (\mu_1 - a_2\Delta t_0) = -a_2 t + \mu'_1 \tag{10}$$

where $\mu'_1 = \mu_1 - a_2\Delta t_0$ and $t$ here are calculated from the start time of cashless activities in a country.

Our next question is what single function for $\ln(1/p - 1)$ that satisfies Eq. (9) for small $t$ and Eq. (8) for large $t$. Let us propose a trial function as follows

$$\ln\left(\frac{1}{p} - 1\right) = (\alpha t + \beta)\left(\gamma - \tanh\left(\frac{t}{T}\right)\right) \tag{11}$$

where $\alpha, \beta, \gamma$, and $T$ are the parameters to be determined. If $t/T \gg 1$, we get

$$\ln\left(\frac{1}{p} - 1\right) \approx (\alpha t + \beta)(\gamma - 1) = (\gamma - 1)\alpha t + (\gamma - 1)\beta \tag{12}$$

which is similar to Eq. (8). If $t/T \ll 1$, we get

$$\ln\left(\frac{1}{p} - 1\right) \approx (\alpha t + \beta)\left(\gamma - \frac{t}{T}\right) = -\frac{\alpha}{T}t^2 + \left(\alpha\gamma - \frac{\beta}{T}\right)t + \beta\gamma \tag{13}$$

which is similar to Eq. (9). Since the slope in **Fig. 6** is negative, we conclude from Eq. (12) that $(\gamma - 1)\alpha < 0$. Referring to **Fig. 7** and considering $T > 0$ we conclude from Eq. (13) that $\alpha > 0$ which implies $\gamma < 1$.

By comparing Eqs. (9) and (13) we conclude that

$$\frac{\alpha}{T} = \frac{b}{2} \tag{14}$$

$$\alpha\gamma - \frac{\beta}{T} = -a_{20} \tag{15}$$

$$\beta\gamma = \mu_2 \tag{16}$$

Then, by comparing equations (8) and (12) we conclude that

$$(\gamma - 1)\alpha = -a_2 \tag{17}$$

$$(\gamma - 1)\beta = \mu'_1 = \mu_1 - a_2\Delta t_0 \tag{18}$$

Equation (14) gives $\alpha = bT/2$. Substituting Eq. (16) into (15) we get a quadratic equation in $\beta$, namely $\beta^2/T - a_{20}\beta - \alpha\mu_2 = 0$. From the solution for $\beta$ we get the value of $\gamma$ from Eq. (16).

In **Fig. 6**, we have obtained the fitting equations for countries with a large CTI and in **Fig. 7** we have obtained the fitting equations for countries with small CTI. An interesting question is, what is the shape of the curves for the countries in **Fig. 6** when it is extended to the inception or transitioning phase and what is the shape of the curves for the countries in **Fig. 7** when it is extended to the tipping point or nearly cashless phase. These equations are generally in the form of Eq. (11), but we must estimate the parameters contained. From the fitting coefficients of **Fig. 6** or **7** we will determine the parameters $\alpha$, $\beta$, and $\gamma$.

We estimate the parameter $\Delta t_0$ as follwos. Equation (2) produces a percolation threshold at $p_c = 1/(z-1)$ [20]. For square cells, $z = 4$ and $p_c = 1/(z-1) = 1/3$. We define $\tau$ as the characteristic width of the curve $p(t)$. Because $\ln(1/p - 1)$ has a different function for conditions when $p(t)$ is small and when $p(t)$ is close to unity, the shape of the curve $p(t)$ is not inversely symmetrical about any point. For countries with small cashless shares ($p < 1/2$), we assume that $\tau$ is the width of the characteristic curve measured from the bottom, taking the distance between $t$ which gives $p(t) = p_c$ and $t$ which gives $p(t) = 1/2$ is the same as $\tau/2$. For countries with large cashless shares ($p > 1/2$), we assume that $\tau$ is the width of the characteristic curve measured from above, taking the distance between $t$ which gives $p(t) = 1/2$ and $t$ which gives $p(t) = 1 - p_c$ is equal to $\tau/2$. For clarity, look at **Fig. 8**.

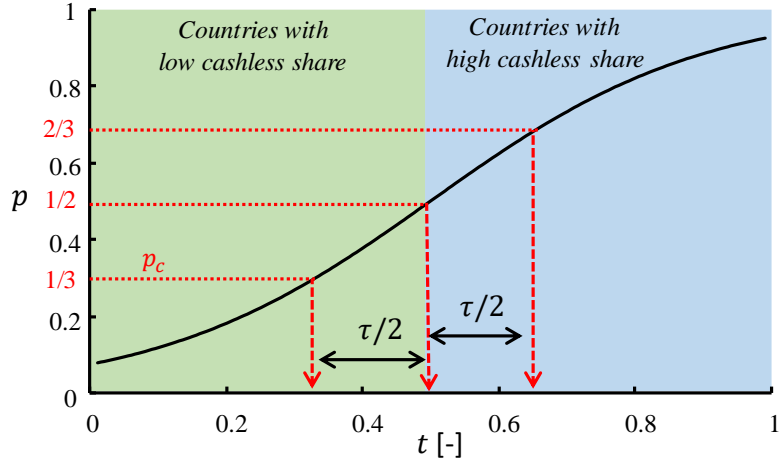

**Figure 8** The definition of parameter $\tau$ from the cashless share curve.

For countries with small cashless shares,

$$\frac{\tau}{2} = t_2 - t_1 \tag{19}$$

where $t_1$ and $t_2$ are solutions of the following quadratic equations

$$k_1 t_1^2 + k_2 t_1 + k_3 = \ln\left(\frac{1}{p_c} - 1\right) \tag{20a}$$

$$k_1 t_2^2 + k_2 t_2 + k_3 = \ln\left(\frac{1}{1/2} - 1\right) = 0 \tag{20b}$$

and $k_1$, $k_2$, and $k_3$ are the coefficients on each curve in **Fig. 7**. On the other hand, for countries with small cashless shares,

$$\frac{\tau}{2} = t_2 - t_1 \tag{21}$$

where, $t_1$ and $t_2$ are solutions of the following linear equations

$$h_1 t_1 + h_2 = \ln\left(\frac{1}{1/2} - 1\right) = 0 \tag{22a}$$

$$h_1 t_2 + h_2 = \ln\left(\frac{1}{1-p_c} - 1\right) \tag{22b}$$

and $h_1$ and $h_2$ are the coefficients on each curve in **Fig. 6**.

As illustrated in **Fig. 9**, we define $\Delta t_0$ as the distance between the initial years in the data by extrapolating the current year that is $\tau$ behind the year corresponding to $p = 1/2$. Based on Eq. (9) we determine $t_{1/2}$ when $p = 1/2$. If $t_{1/2} < 0$, we have $\Delta t_0 = |t_{1/2}| + \tau$. On the other hand, if $t_{1/2} > 0$, we have $\Delta t_0 = \tau - t_{1/2}$. The results of the calculation of $\Delta t_0$ for the countries in **Fig. 6** are shown in **Table 1**. As parameter $T$ we have selected 50 years for all countries. This value is still speculation, but results in a match between the data and the results of the simulation.

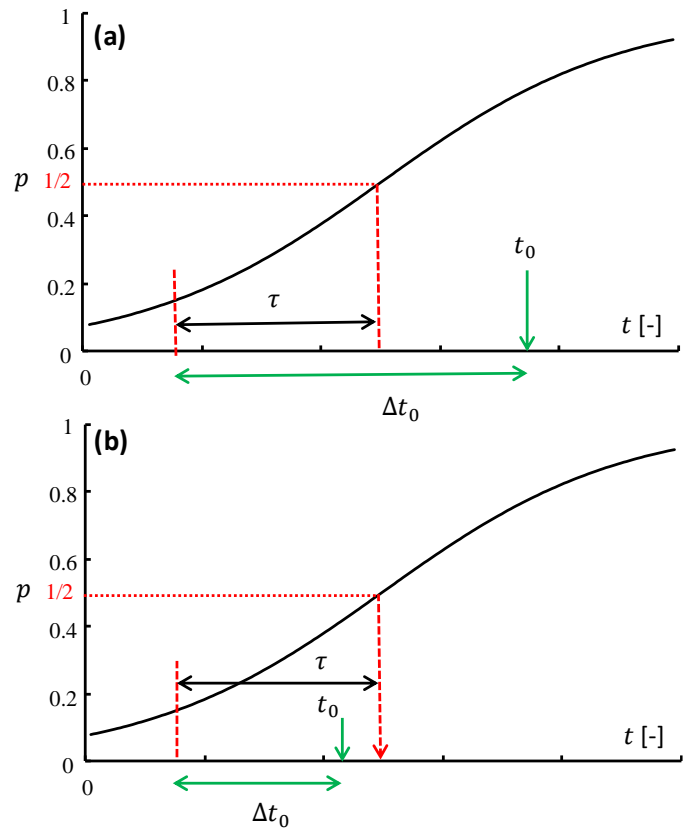

**Figure 9** The definition of the parameter $\Delta t_0$ of the cashless share curve for the countries in **Fig. 6**: (a) if $p(t_0) > 1/2$ and (b) if $p(t_0) < 1/2$.

**Table 1** Parameters α, β, γ, T, and $\Delta t$ for the countries studied.

| Country | α [year$^{-1}$] | β | γ | T [year] | $\Delta t_0$ [year] |
|---|---|---|---|---|---|
| Hungary | 0.160 | 3.498 | 0.397 | 50 | 0 |
| Italy | 0.305 | 5.195 | 0.228 | 50 | 0 |
| Japan | 0.325 | 6.697 | 0.269 | 50 | 0 |
| Croatia | 0.585 | 7.857 | 0.206 | 50 | 0 |
| Slovenia | 0.165 | 5.097 | 0.190 | 50 | 0 |
| Slovakia | 0.160 | 2.436 | 0.397 | 50 | 0 |
| Finland | 0.119 | 1.679 | 0.30 | 50 | 15 |
| Sweden | 0.153 | 1.583 | 0.21 | 50 | 11 |
| UK | 0.133 | 1.920 | 0.30 | 50 | 15 |
| Netherland | 0.122 | 1.411 | 0.26 | 50 | 13 |
| Denmark | 0.261 | 4.367 | 0.30 | 50 | 16 |
| Portugal | 0.155 | 2.054 | 0.30 | 50 | 13 |

To estimate the parameters for the country shown in **Fig. 6**, we use Eqs. (17) and (18). But the problem here is that we have three parameters to define, $\alpha$, $\beta$, and $\gamma$, whereas we only have two equations. Therefore, one parameter should be chosen freely. If we look at **Table 1**, the values of the parameters $\alpha$ and $\beta$ for the countries in **Fig. 7** (Hungary, Italy, Japan, Croatia, Slovenia, and Slovakia), vary quite sharply, while the parameter $\gamma$ varies in a fairly small range between 0.2 – 0.4. If we assume that the value of $\gamma$ is close to universal around 0.3 then we can estimate the parameters $\alpha$ and $\beta$ in Eqs. (17) and (18). We get the estimated parameters in **Table 1**.

**Figure 10** shows the comparison of data [34- 37] with curve fitting Eq. (11) using the parameters in **Table 1** for the group of countries in **Fig. 6**. We see a good match between the data and the fitting results which leads to the conclusion that, in general, all countries experience changes in $\ln(1/p - 1)$ as a function of time according to Eq. (11). In the early stages of cashless activities (inception or transitioning), $\ln(1/p - 1)$ changes quadratic to time, while when entering the tipping point or nearly cashless phase, $\ln(1/p - 1)$ decreases linearly with time. When $t \to \infty$, we will get $\ln(1/p - 1) \to -\infty$ which means $p \to 1$, as expected.

**Figure 11** shows a comparison of the data [28, 30-32] with the curve fitting Eq. (11) using the parameters in **Table 1** for the group of countries in **Fig. 7**. We see a good match between the data and the fitting results which leads to the conclusion that, in general, all countries experience a change in $\ln(1/p - 1)$ as a function of time according to Eq. (11).

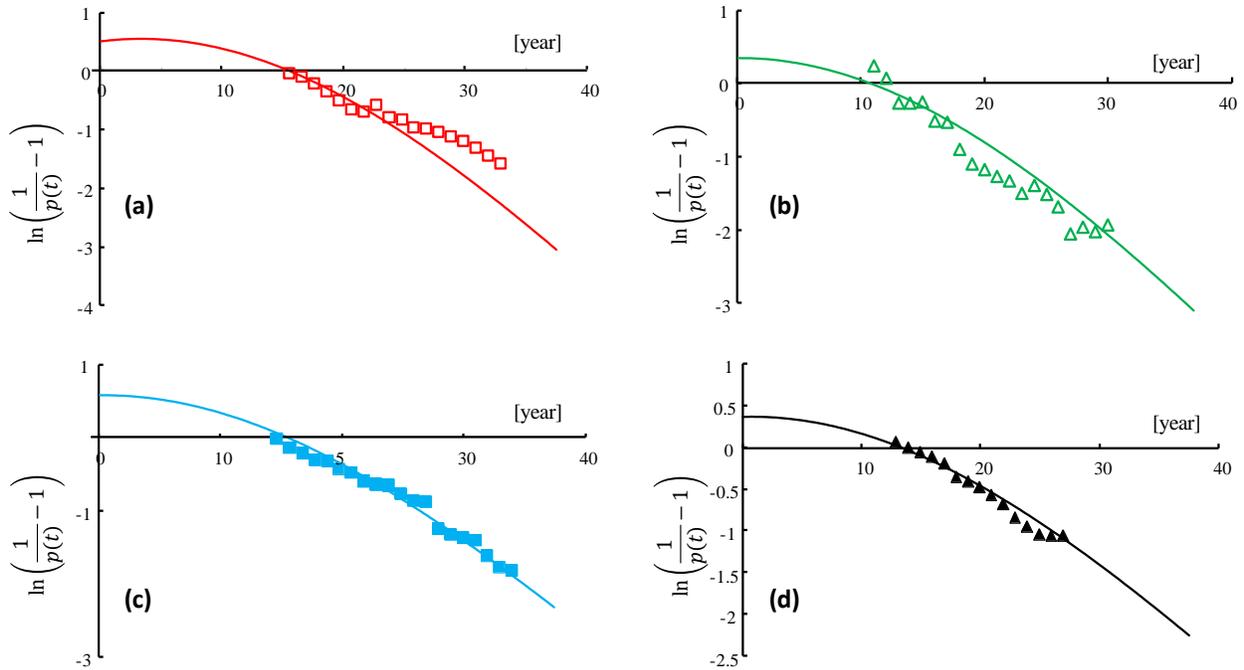

**Figure 10** Comparison between data (symbols) and fitting results (curves) using Eq. (11) and parameters in Table 1 for the countries: (a) Finland, (b) Sweden (c) UK, and (d) Netherland. Data obtained from (a): [35] (b): [37] (c): [36] and (d): [34]

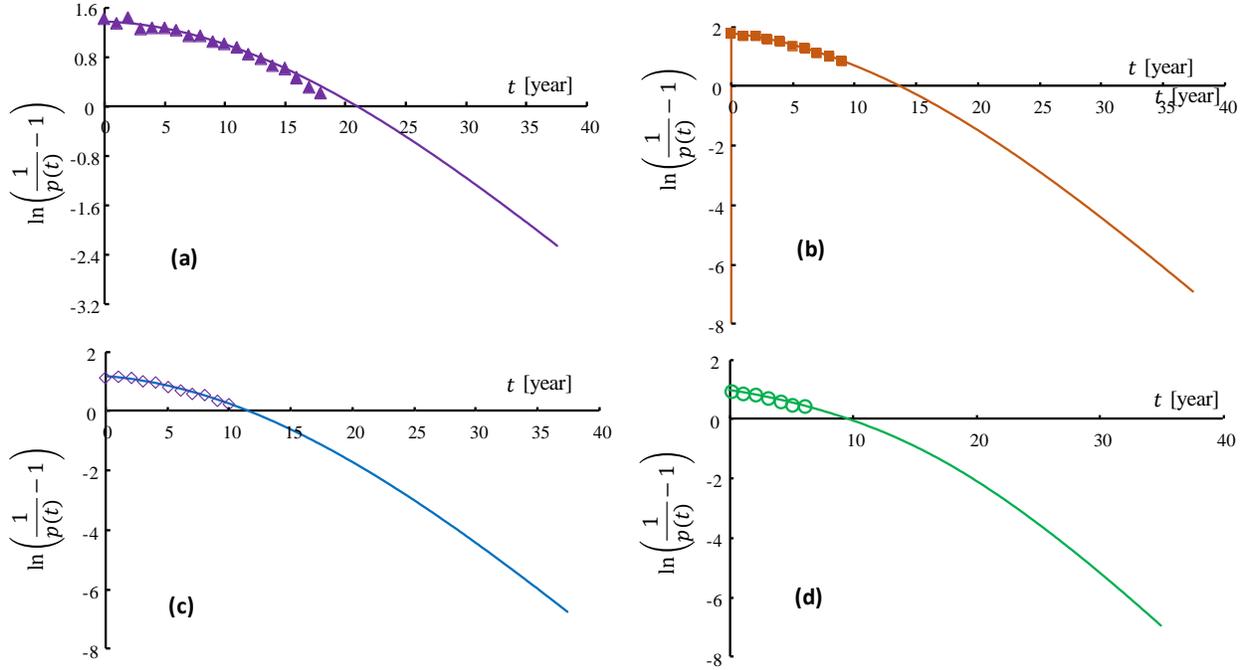

**Figure 11** Comparison between data (symbols) and fitting results (curves) using Eq. (11) and parameters in **Table 1** for countries: (a) Hungary, (b) Italy, (c) Japan, and (d) Slovenia. Data obtained from (a): [32 (b): [31] (c): [28] (d): [30]

The slope of Eq. (11) is

$$a_2 = -\frac{d}{dt}\ln\left(\frac{1}{p} - 1\right) = -\alpha\left(\gamma - \tanh\left(\frac{t}{T}\right)\right) + \frac{1}{T}(\alpha t + \beta)\operatorname{sech}^2\left(\frac{t}{T}\right) \qquad (23)$$

This function has a peak if the slope at $t = 0$ if $a_2$ is negative. The slope at $t = 0$ is $a_2(0) = -\alpha\gamma + \beta/T$. For example, for Hungary and the Netherlands, we get $a_2(0) = 0.0064$ and $-0.00372$, respectively, which means that for the Hungary there is no peak, while for the Netherland there is a peak. The peak location is at $t$ which satisfies the equation $-(\gamma - \tanh(t/T)) + (t/T + \beta/(T\alpha))\operatorname{sech}^2(t/T) = 0$. For the Netherlands, we find that the peak occurs at $t \approx 0.72$ year.

From Eqs. (11) and (12) we can get the following equation

$$\ln\left(\frac{1}{p(t)} - 1\right) \approx \ln\left(\frac{1}{p(t)} - 1\right)\Big|_{t/T \gg 1} \left(\frac{\gamma - \tanh(t/T)}{\gamma - 1}\right) \qquad (24)$$

It can be seen from Eq. (24) that $\ln(1/p - 1)$ can be seen as the result of "modulation" of similar quantities at large t values with the modulation function $(\gamma - \tanh(t/T))/(\gamma - 1)$. Since both $\gamma$ and $T$ can be considered to be constant for all countries (see **Table 1**) we conclude that the modulating function is nearly universal for all countries. Even if there are variations in $\gamma$ and $T$ the resulting variations are not too large.

From Eq. (24) we also get $\ln(1/p - 1) = 0$ when $\gamma - \tanh(t/T) = 0$ which gives $t = T \operatorname{atanh}(\gamma) \approx 0.31\,T$. This condition also produces $p = 1/2$, which is the transition from the transitioning state to the tipping point state.

### 3.3 CTI grow rate

Next we will calculate the rate of change of CTI with time,

$$\frac{d\,CTI}{dt} = 10 \frac{1}{\sigma_e} \frac{d\sigma_e}{dt} \qquad (25)$$

The solution for $\sigma_e$ is $\sigma_e = \left(-B(p) + \sqrt{B(p)^2 - 4AC}\right)/2A$. Tetapi $d\sigma_e/dt = (d\sigma_e/dp)(dp/dt)$ and

$$\frac{d\sigma_e}{dp} = \frac{1}{2A} \frac{dB}{dp} \left(\frac{B(p)}{\sqrt{B(p)^2 - 4AC}} - 1\right) \qquad (26)$$

so we get the following equation

$$\frac{d\sigma_e}{dt} = \frac{(A+1)\Delta\sigma}{2A} \left(\frac{\left(p - \frac{\sigma_c A - \sigma_{cl}}{(A+1)\Delta\sigma}\right)}{\sqrt{\left(p - \frac{\sigma_c A - \sigma_{cl}}{(A+1)\Delta\sigma}\right)^2 - \frac{4AC}{(1+A)^2\Delta\sigma^2}}} + 1\right) a_2 p(1-p) \qquad (27)$$

where $\Delta\sigma = \sigma_{cl} - \sigma_c$.

**Figure 12** is the result of calculating CTI growth for the six selected countries ((a) Hungary, (b) Italy, (c) Japan, (d) UK, (e) Netherland, (f) Sweden). We have used the parameters in Table 1 as well as $\sigma_c = 1$, $\sigma_{cl} = 10$, $z = 4$, $A = 1$, and $C = -\sigma_c \sigma_{cl} = -10$. For all countries,

CTI growth initially increases with time, peaks at a certain time and then falls to near zero when time is very large. Interesting to see are Italy and Japan. In the early stages, the growth rate of CTI is very large then closes to zero at large times. This means that in the last year, there was no significant growth in the CTI of the two countries due to almost stagnant cashless shares. Hungary, UK, Netherland, and Sweden are still experiencing high growth to a large extent.

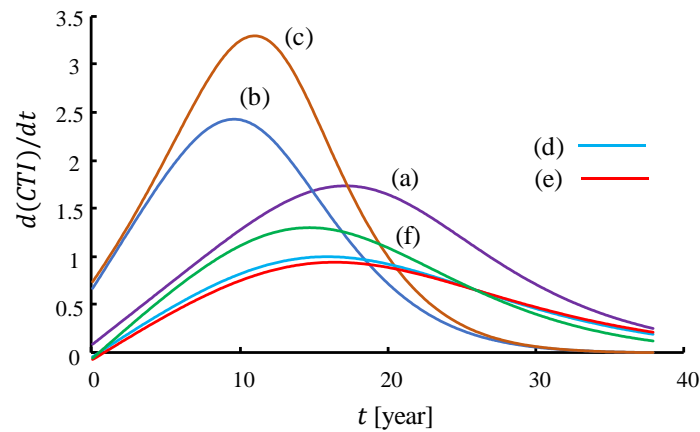

**Figure 12** The rate of change of the CTI as a function of time for countries: (a) Hungary, (b) Italy, (c) Japan, (d) UK, (e) Netherland, (f) Sweden. We have used the parameters in **Table 1**.

### 3.4 Policy effect on the cashless share

The theoretical curves in **Fig. 10** or **11** have been obtained using the parameters in **Table 1**. However, these parameters are obtained from the calculation process on the data available to date. If a country's financial policy does not change, then the curve can predict the CTI for the next few years. However, government policies that issue new financial policies can result in deviations from the forecast curve. Especially for countries with small cashless shares, the government can take policies to accelerate the increase in cashless shares. For example, the declaration of cashless economy policy by the Central Bank of Nigeria in 2012, has improved Nigeria's payment landscape [41]. However, for countries that are approaching a cashless society, there is no need for a meaningful policy to increase cashless share, because the policy was implemented much earlier. For example, in Sweden, in 1990's, the system for card payments was build and checks were phased out via new fees [42]. If the policy taken by the government is to accelerate the growth of the CTI, then the policy means accelerating the growth of cashless share towards unity. From Eq. (11) it appears that if $p$ goes to unity faster then the right side will go to negative infity faster. This happens when $\alpha$ gets bigger. Thus, the policy that accelerates the increase in cashless share is manifested in Eq. (11) as an increase in $\alpha$.

Let us assume that the policy is interpreted as assigning a time factor to the parameter $\alpha$. To make the data in **Fig. 11** for countries with low cashless shares still fit with a curve that uses the parameter $\alpha$ as a function of time, these parameters must change slowly with time when time is small and only change significantly when time is very large. One of the possible equations is the sigmoid form, and one of them is

$$\alpha'(t) = \alpha + \epsilon \left(1 + \frac{2}{\pi} \arctan\left(\frac{t-t_i}{\omega}\right)\right) \tag{28}$$

where $\epsilon$ is the policy impact parameter, $t_i$ is the start time for the policy to be implemented, and $\omega$ is a parameter that measures how quickly the impact of the policy will increase cashless share (the smaller $\omega$, the faster the impact of the policy). Equation (28) is not necessarily an accurate function. Selection of other functions is still possible. A more representative function can be obtained by analyzing changes in the cashless share of a country that has just implemented a policy to increase its cashless share. From the data on the share of several years before and after the policy was implemented, we can estimate the change in $\alpha$ more accurately. It is clear from the above equation that, if $t < t_i$ and $|(t-t_i)/\omega| \gg 1$, $\alpha'(t) \to \alpha$, and, if $t > t_i$ ann $(t-t_i)/\omega \gg 1$, $\alpha'(t) \to \alpha + \epsilon$. If the policy is carried out several times, Eq. (28) takes the form

$$\alpha'(t) = \alpha + \sum_i \epsilon_i \left(1 + \frac{2}{\pi} \arctan\left(\frac{t-t_i}{\omega_i}\right)\right) \tag{29}$$

where $\epsilon_i$ is the impact of i-th policy and $\omega_i$ measures how fast the impact of the i-th policy on increasing the cashless share.

**Figure 12** is an example for Japan if the policy is taken at $t_i = 15$ years assuming $\omega = 10$ years. The choice of $\omega = 10$ years is still speculation, but based on information reported by Equinix.com [40] that the Japanese government plans to double its cashless share from 20% in 2017 to 40% in 2027 (10 years span). So, the impact of implementing the policy occurs in a span of about 10 years. We use three values of $\epsilon = 0.0$, 0.1, and 0.2. With increasing $\epsilon$, $\ln(1/p - 1)$ decreased more rapidly with increasing time and the CTI increased more rapidly with increasing time (**Fig. 12**(a)). It is clear from **Figs. 12(a)** and **(b)** that the change occurs at $t > t_i = 15$ years.

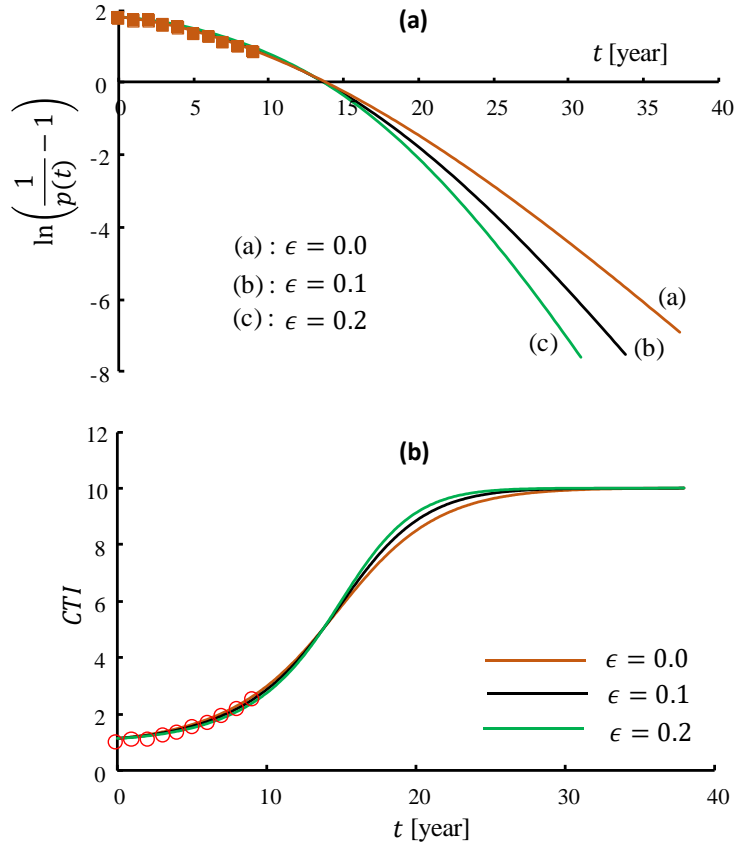

**Figure 12** Effects of policy on changes in: (a) $\ln(1/p - 1)$ and (b) $CTI$ of Japan using equation (28). In the calculation we have used the parameters in **Table 1**, $t_i = 15$ years, and $\omega = 10$ years. Three parameter values $\epsilon = 0.0$, 0.1, and 0.2 were used. The symbols are the result of data processing from [28] and the curves are the calculation result.

## 4. Conclusion

The cashless transaction index (CTI) can become a representative parameter to grade the cashless behavior of a country. We have shown that the Nordic countries have high CTIs, in accordance with the condition of these countries which are soon entering a cashless society. This proves that the EMA method is representative enough to determine cashless grading of a country by analogizing individuals who carry out cashless transactions as conductor elements in the composite and individuals who carry out cash transactions as insulator elements in the composite. The introduced time dependent cashless share equation can explain well the changes in the cashless share of selected countries over a period of up to about 20 years.